\documentclass[apj]{emulateapj}
\usepackage{apjfonts}

\usepackage{amsmath}
\usepackage{natbib}

\submitted{ApJ, in press}

\usepackage{graphicx}
\usepackage[figuresright]{rotating} 
\usepackage{float}       
\usepackage{afterpage}   

\newcommand {\IN}{$\Gamma$}
\newcommand {\SO}{4U~1543$-$47}
\newcommand{\hexte}{HEXTE}
\newcommand{\wsim}{\ensuremath{\sim}}
\newcommand{\rxte}{\emph{RXTE}}
\newcommand{\msun}{$\rm M_{\odot}$}
\newcommand{\tin}{$T_{in}$}

\begin{document}


\title{Multi-wavelength observations of the Galactic black hole transient
4U~1543$-$47 during outburst decay: state transitions and jet contribution}

\author{E. Kalemci\altaffilmark{1},
        J. A. Tomsick\altaffilmark{2},
        M. M. Buxton\altaffilmark{3},
        R. E. Rothschild\altaffilmark{2},
        K. Pottschmidt\altaffilmark{4,5},
        S. Corbel\altaffilmark{6},
        C. Brocksopp\altaffilmark{7},
        P. Kaaret\altaffilmark{8}
}

\altaffiltext{1}{Space Sciences Laboratory, 7 Gauss Way, University of
California, Berkeley, CA, 94720-7450, USA}

\altaffiltext{2}{Center for Astrophysics and Space Sciences, Code
0424, University of California at San Diego, La Jolla, CA,
92093-0424, USA}

\altaffiltext{3}{Department of Astronomy, Yale University, New Haven, CT, 06511, USA }

\altaffiltext{4}{Max-Planck-Institut f\"ur Extraterrestrische Physik, Giessenbachstr. 1, 85748 Garching, Germany}

\altaffiltext{5}{\emph{INTEGRAL} Science Data Centre, Chemin d'\'Ecogia 16, 1290 Versoix, Switzerland}

\altaffiltext{6}{Universit\'e Paris VII and Service d'Astrophysique, CEA Saclay, F-91191 Gif sur Yvette, France}

\altaffiltext{7}{Mullard Space Sciences Laboratory, Holmbury St. Mary, Dorking, Surrey RH5 6NT}

\altaffiltext{8}{Department of Physics and Astronomy, University
of Iowa, Iowa City, Iowa, 52242, USA}


\begin{abstract}

Multiwavelength observations of Galactic black hole (GBH) transients during the
state transitions and in the low/hard state may provide detailed information on
 the accretion structure of these systems. \SO\ is a GBH transient that was
covered exceptionally well in X-ray and infrared (daily
observations) and reasonably well in optical and radio during its
outburst decay in 2002. When all the available information is
gathered in the intermediate and the low/hard state, \SO\ makes an
important contribution to our understanding of state transitions
and the role of outflows on the high energy emission properties of
black hole binaries. The evolution of the X-ray spectral and
temporal properties and the IR light curve place strong
constraints on different models for explaining the overall
emission from accreting black holes. The overall spectral energy
distribution is consistent with synchrotron origin for the optical
and infrared emission, however, the X-ray flux is above the
power-law continuation of the optical and infrared flux. The
infrared light curve, the \hexte\ light curve and the evolution of
the X-ray photon index indicate that the major source of hard
X-rays cannot be direct synchrotron radiation from an acceleration
region in a jet for most of the outburst decay.

\end{abstract}




\section{Introduction}\label{sec:intro}

Galactic black hole (GBH) transients are binary systems that can show orders
of magnitude increases in luminosity during outbursts caused by sudden,
dramatic increases of mass accretion onto compact objects.
 During an outburst, a GBH transient often follows a specific sequence of
X-ray spectral states; it is in the ``hard'' or ``low-hard'' state (LH)
at the beginning, and also during the decay of an outburst. In between the 
rise and the decay, when the 2--10 keV luminosity is high, the system is
usually in the ``thermal dominant'' state (TD, historically
this state was called high/soft state), and sometimes in the ``steep power-law
state \citep[historically very-high state, see][for detailed
discussion of spectral states]{McClintock03}. They can also be found in
intermediate states (IS), where source characteristics do not fit into
the TD, or LH states, but show various combinations of these states,
especially close to state transitions \citep{McClintock03}. The sequence of
spectral states might be complicated, or very simple as some sources stay
in the LH state throughout their outbursts \citep[][and references
therein]{Brocksopp04,Kalemci_tez}. However, they are almost always found
in the LH state during outburst decay \citep{Kalemci_tez}, before they
return to their quiescent states (perhaps a continuation of the LH
state at very low luminosities, \citealt{Tomsick04,Kong02,Corbel00}).

In terms of timing, the TD state is characterized by a lack of, or a very low
level of variability (a few \% rms in 0.04--4 Hz), whereas the LH state shows
strong variability (\wsim 30\% rms in the same band), along with
quasi-periodic oscillations (QPOs) in the power-density spectrum (PDS). Other
Fourier-analysis related timing tools such as ``coherence'' and ``lag''
behavior are also different in each state (see \citealt{Vaughan97,Nowak99} for 
detailed definitions of these quantities). The mean coherence is very
high (\wsim1) in the LH state and the mean lag is either zero or very low
between the 3--6 keV and 6--15 keV bands in the 1--10 Hz band. During the IS
and during transitions, however, the coherence is lower and the lag is higher
in the same frequency band and between the same energy bands 
\citep{Kalemci_tez,Pottschmidt00}.

The relation between the radio emission and the X-ray states has been well 
established (\citealt{McClintock03}, see also \citealt{Corbel04}, and a recent
review by \citealt{Fender03b}). In the TD state, the radio
emission is often quenched \citep{Fender99,Corbel00}. Optically
thin outflows are sometimes detected during state transitions
\citep{Fender01_c,Corbel01}, and powerful, compact jets are always
observed in the LH state \citep{Fender01b}. In addition, there exists a global 
correlation between the X-ray flux and the radio flux for different sources at 
different luminosities in the LH state \citep{Corbel03,Gallo03}. The base of 
the jet might provide the energetic electrons that would create the X-ray 
power-law in the spectrum, establishing the link between the radio and X-ray 
flux \citep{Fender01b}. The fact that the lags are higher during state
transitions, when the optically thin, large  outflows are observed, may also 
point out a relation between the radio jet  and energetic electrons 
\citep{Kalemci02,Pottschmidt00}. An alternative model is that the X-ray 
emission is synchrotron in nature and comes directly from a shock region in 
the jet \citep{Falcke99,Markoff01}. This  model can explain the radio - X-ray 
flux correlation naturally; however, it lacks any prediction on the details of 
X-ray spectrum \footnote{Only recently, the reflection from the disk has been 
incorporated into the synchrotron jet model by \cite{Markoff04}. See also 
\cite{Kording04} on a recent work on timing properties from this model.}.

The daily observations of the transients in the optical and the
infrared (OIR) by the YALO consortium \citep{Bailyn99} provided
another dimension in the study of GBHTs. A secondary maximum in $V$, $B$, $I$, 
$J$, and $K$ bands has been observed during the outburst decay of \SO\ 
\citep{Buxton04}. A similar secondary maximum has also been observed in $V$, 
$I$ and $H$ bands during the outburst decay of XTE~J1550$-$564 in 2000 
\citep{Jain01_b}. The properties of the optical emission during the secondary 
maximum of the 2000 outburst of XTE~J1550$-$564 indicate a synchrotron origin
from a jet \citep{Corbel01}, rather than X-ray reprocessing at the
outer parts of the accretion disk, which was suggested to explain
the optical light curves of the same source in the 1998 outburst
\citep[and also other GBH transients,][]{Jain01_b,Jain01}. The
X-ray reprocessing for the origin of optical emission was also
questioned by \cite{Brocksopp04} because of a lack of correlation
between X-ray and the optical light curves of V404 Cyg,
GRO~J1719$-$24, GRO~J0422$+$32, GS~1354$-$64 and XTE~J1118$+$480.
The observations of GX~339$-$4 in the optical and the near
infra-red (nIR) in the LH state showed a non-thermal optically
thin synchrotron component that extrapolates down to the X-ray
spectrum \citep{Corbel02}. A very recent study on GX~339$-$4 with
good optical, infrared and X-ray monitoring campaign points out
that the nIR has a synchrotron origin from a jet in the LH state,
however the origin of optical emission may be a combination of
jet, disk and possibly a compact corona for this source
\citep{Homan04}. In addition, the LH state during the rise of the
2002 outburst of GX 339$-$4 showed a strong correlation between
X-ray flux and optical / nIR, similar to the radio - X-ray flux
correlation \citep{Homan04}.

One of the most interesting epochs of an outburst is its decay,
because it is almost guaranteed that there will be a transition to
the LH state which provides additional information about the
system through various timing analysis techniques, and also
through strong radio emission. Analysis of state
transitions helps us probe the accretion dynamics of these systems.
Our group has been observing GBH transients during outburst decays
in X-rays with the \emph{Rossi X-ray Timing Explorer} (\rxte) and
in radio to understand the evolution before, during and after the
transition to the LH \citep{Kalemci01,Tomsick01b,Kalemci02,Tomsick03_2}. A 
uniform analysis of all GBH transients observed with approximately daily 
coverage with \rxte\ between 1996 and 2001 resulted in a better understanding 
of the evolution of spectral and temporal parameters during the outburst decay 
\citep{Kalemci03}. The most striking of all results is that the sharpest 
change indicating a state transition is observed in the timing 
properties (usually a jump in the rms amplitude of variability from less than 
a few percent to more than tens of percent in less than a day). This change in
the rms amplitude is often (but not always) accompanied by a sharp
increase in the power-law flux.  This sharp change in rms
amplitude of variability is noted as the time of state transition
in \cite{Kalemci03}, and the same will be applied here. 

\begin{figure}[t]
\plotone{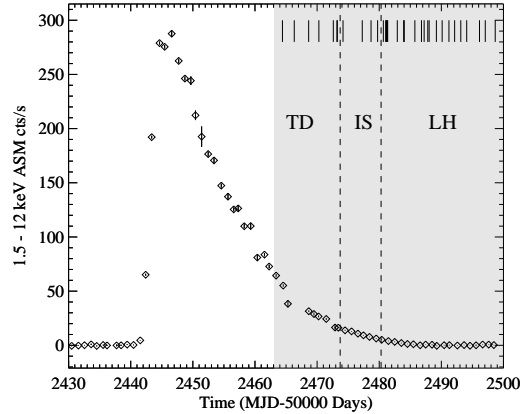}
\caption{\label{fig:asm}
The 1.5--12 keV ASM light curve of the 2002 outburst of \SO. We worked on the
decay part of the outburst, and analyzed \rxte\ data shown in the gray region.
The vertical short lines indicate pointed PCA observations. The dashed lines
indicate approximate times of state transitions.
}
\end{figure}

In this work, we will try to obtain a coherent picture of changes in physical
properties of a GBH transient, \SO, around state transitions and also deep in
the LH state using multi-wavelength observations. \SO\
was discovered by the \emph{Uhuru} satellite on August 17, 1971
\citep{Matilsky72}. The source was observed again in outburst, in 1983, 1992,
and also in 2002 \citep{Kitamoto84,Harmon92_iauc,Brocksopp04,Miller02_iauc}.
The optical counterpart was found by \cite{Pedersen83}. Initial dynamical
mass measurements during quiescence established this source as a black hole
binary system with a compact object mass between 2.7 and 7.5 \msun\
\citep{Orosz98}. A more accurate value of $\rm 9.4\pm2.0$ \msun\ is given
in \cite{Park03} based on a work in preparation by Orosz et al. The 2002 
outburst was first detected by the All Sky Monitor (ASM) on \rxte\ on
MJD 52442 \citep{Miller02_iauc,Park03}. Around a month later, as the outburst
was decaying, our group started its daily monitoring campaign with \rxte,
and caught the transition to a harder state on MJD 52479
\citep{Kalemci02_atel}, although subsequent analysis in this work
indicates that the transition started earlier. The analysis of \rxte\ data 
before the LH state, with an emphasis on broad iron reflection lines, is given 
by \cite{Park03}. Fig.~\ref{fig:asm} shows the ASM light curve of the overall 
outburst, the epoch we analyzed, and the dates of pointed observations used in 
this work. The source was also detected and observed in radio and optical 
bands \citep[see][$\S$~\ref{subsec:radio}, and $\S$~\ref{subsec:opir} in this 
work for more details]{Park03,Buxton04}. Here, we report on the \rxte\
observations during outburst decay, and combine the results with
the optical, IR and radio information to understand jet formation
and its effects on spectral and temporal properties of \SO.

\section{Observations and Analysis}\label{sec:obs}

\subsection{The RXTE observations}\label{subsec:rxteobs}

We triggered our monitoring program (P70124, PI Tomsick) with \rxte\ after the
source's ASM count rate dropped below 15 cts/s, and the first observation took
place on MJD~52478.7. The source was already in transition from the TD to the
LH state, showing broad band variability and QPOs \citep{Kalemci02_atel}. While
 the \wsim daily 1-3 ks monitoring from the program P70124 was going on, we
triggered P70128 (PI Kalemci) which provided longer exposures (\wsim 20 ks) to
investigate spectral and temporal properties of the source in greater detail
than the daily monitoring observations. For this work, we used the long 
observations, and the first 18 observations of the monitoring program that 
provide high quality spectral information. A very interesting part of the 
outburst, the onset of state transition, occurred before our monitoring 
observations had begun. To characterize this part, we used archival \rxte\ 
observations from P70133 (PI McClintock). Table~\ref{table:spe_par} shows the 
list of the observations that we used in 
this work.

\begin{table*}[ht]
\caption{\label{table:spe_par} Spectral Parameters}
\scriptsize
\begin{tabular}{r|c|c|c|c|c|l} \hline \hline
Obs.\#\footnote{A for observations from P70133 (PI McClintock), B for observations from P70124 (PI Tomsick), and C for observations from P70128 (PI Kalemci). Observation 14 from P70128 is divided into four.} & Date & \IN\ & \tin\ & Power-law flux\footnote{in units of $\rm 10^{-10} \; ergs \; cm^{-2} \; s^{-1}$} & DBB flux\footnote{in units of $\rm 10^{-10} \; ergs \; cm^{-2} \; s^{-1}$} & Notes \\   1A & 52464.42 &  $2.32 \pm 0.09$ & $0.70 \pm 0.01$ & $10.19 \pm 0.97$ & $53.92 \pm 0.47$ \\
   2A & 52466.32 &  $2.39 \pm 0.07$ & $0.67 \pm 0.01$ & $11.96 \pm 0.91$ & $38.20 \pm 0.43$ \\
   3A & 52468.64 &  $2.10 \pm 0.11$ & $0.65 \pm 0.01$ & $5.06 \pm 0.68$ &
$29.44 \pm 0.28$ \\
   4A & 52470.28 &  $2.32 \pm 0.04$ & $0.62 \pm 0.01$ & $4.99 \pm 0.59$ &
$21.72 \pm 0.24$ \\
   5A & 52472.61 &  $2.49 \pm 0.15$ & $0.57 \pm 0.01$ & $4.69 \pm 0.50$ &
$12.84 \pm 0.14$ \\
   6A & 52473.19 &  $2.57 \pm 0.03$ & $0.54 \pm 0.01$ & $7.69 \pm 0.58$ &
$10.62 \pm 0.14$ \\
   7A & 52473.25 &  $2.51 \pm 0.04$ & $0.54 \pm 0.01$ & $6.36 \pm 0.52$ &
$10.99 \pm 0.16$ \\  \hline
   8A & 52474.17 &  $2.54 \pm 0.03$ & $0.53 \pm 0.01$ & $10.38 \pm 0.29$ & $7.781 \pm 0.103$  & Variability, state transition \footnote{The line above
indicates a transition to the IS from TD state}\\
   9A & 52477.27 &  $2.43 \pm 0.01$ & $0.47 \pm 0.01$ & $12.78 \pm 0.27$ & $2.783 \pm 0.074$ & QPO \\
  10B & 52478.68 &  $2.43 \pm 0.01$ & $0.43 \pm 0.01$ & $11.75 \pm 0.32$ & $1.676 \pm 0.058$ & QPO \\
  11B & 52479.74 &  $2.40 \pm 0.02$ & $0.42 \pm 0.02$ & $9.26 \pm 0.12$ &
$1.324 \pm 0.041$ & QPO \\ \hline
  12B & 52480.66 &  $2.10 \pm 0.02$ & $0.37 \pm 0.03$ & $12.29 \pm 0.10$ &
$0.365 \pm 0.026$ & Spectral hardening, QPO  \footnote{The line above indicates
a transition from IS to LH state.}\\
  13C & 52481.03 &  $2.10 \pm 0.01$ & $0.33 \pm 0.03$ & $11.44 \pm 0.11$ & $0.305 \pm 0.018$ & High energy cut-off, QPO \\
14C.a & 52481.16 &  $2.02 \pm 0.01$ & $0.40 \pm 0.02$ & $11.56 \pm 0.09$ & $0.360 \pm 0.028$ & High energy cut-off, QPO \\
14C.b & 52481.23 &  $2.00 \pm 0.01$ & $0.35 \pm 0.03$ & $12.01 \pm 0.11$ & $0.237 \pm 0.019$ & High energy cut-off, QPO \\
14C.c & 52481.29 &  $1.96 \pm 0.01$ & $0.33 \pm 0.01$ & $12.20 \pm 0.11$ & $0.205 \pm 0.012$ & High energy cut-off, QPO \\
14C.d & 52481.36 &  $1.97 \pm 0.01$ & $0.35 \pm 0.02$ & $11.97 \pm 0.21$ & $0.217 \pm 0.018$ & High energy cut-off ?, QPO \\
  15B & 52482.91 &  $1.78 \pm 0.04$ & 0.35 & $9.77 \pm 0.09$ & $0.116 \pm 0.036$ & Smedge optical depth begins to drop, the disk temperature is fixed from this observation on.\\
  16B & 52483.93 &  $1.70 \pm 0.02$ & 0.35 & $8.00 \pm 0.09$ & $0.056 \pm 0.015$ & IR flux begins to rise \\
  17B & 52484.03 &  $1.71 \pm 0.02$ & 0.35 & $7.88 \pm 0.16$ & $0.031 \pm 0.016$ \\
  18B & 52485.74 &  $1.65 \pm 0.04$ & 0.35 & $4.49 \pm 0.06$ & $0.024 \pm 0.016$ \\
  19B & 52486.80 &  $1.64 \pm 0.04$ & 0.35 & $3.17 \pm 0.03$ & $0.027 \pm 0.008$ \\
  20C & 52487.23 &  $1.63 \pm 0.01$ & 0.35 & $2.62 \pm 0.01$ & $0.028 \pm 0.004$ & MOST radio detection.\\
  21C & 52487.82 &  $1.70 \pm 0.02$ & 0.35 & $2.17 \pm 0.03$ & $0.023 \pm 0.002$ \\
  22B & 52488.03 &  $1.68 \pm 0.02$ & 0.35 & $1.96 \pm 0.03$ & $0.020 \pm 0.004$ & Smedge is not required, Galactic ridge emission ($<$5\% of the total flux) is included\\
  23B & 52489.20 &  $1.70 \pm 0.04$ & 0.35 & $1.25 \pm 0.03$ & $0.020 \pm 0.005$ \\
  24B & 52490.13 &  $1.75 \pm 0.02$ & 0.35 & $1.05 \pm 0.01$ & $0.020 \pm 0.005$ &  Only PCA used from this obs on. ATCA radio detection.\\
  25B & 52491.22 &  $1.88 \pm 0.04$ & 0.35 & $0.76 \pm 0.02$ & $0.014 \pm 0.001$ \\
  26B & 52492.14 &  $2.06 \pm 0.07$ & 0.35 & $0.56 \pm 0.02$ & $0.011 \pm 0.002$ & No timing after this observation, poor statistics.\\
  27B & 52493.13 &  $1.99 \pm 0.06$ & 0.35 & $0.47 \pm 0.02$ & $0.013 \pm 0.002$ & Galactic ridge emission $<$15\% of the total flux. \\
  28B & 52494.08 &  $2.10 \pm 0.05$ & 0.35 & $0.47 \pm 0.02$ & $0.013 \pm 0.002$ \\
  29B & 52496.14 &  $2.24 \pm 0.07$ & 0.35 & $0.30 \pm 0.01$ & $0.010 \pm 0.001$ & Galactic ridge emission $<$25\% of the total flux. \\
  30B & 52497.07 &  $2.11 \pm 0.10$ & 0.35 & $0.29 \pm 0.01$ & $0.010 \pm 0.002$ \\
  31B & 52498.67 &  $2.22 \pm 0.12$ & 0.35 & $0.22 \pm 0.01$ & $0.008 \pm 0.002$ \\
\end{tabular}
\end{table*}

\subsubsection{X-ray spectral analysis}\label{subsubsec:rxtespec}

For the major part of the outburst, we used both the PCA and the HEXTE
instruments on \rxte\ for the spectral analysis (see \citealt{Bradt93} for
instrument descriptions). For the PCA, the 3--25 keV band
was used, and the response matrix and the background model were created
using the standard FTOOLS (version 5.3) programs. We added 0.8\% up to 7 keV,
 and 0.4\% above 7 keV as systematic error based on the fits to Crab
observations close to our observations \citep[for the details of
how we estimated systematic uncertainties, see][]{Tomsick01b}. We used all
available PCUs for each observation, choosing the combination that would
provide the maximum number of counts.

The 15--200 keV band was used for the HEXTE data. We used the response created
by the FTOOLS, and applied the necessary deadtime correction
\citep{Rothschild98}. The HEXTE background subtraction is performed by
alternating between observations of source and background fields.
The relative normalization between the PCA and the HEXTE is kept free. The
HEXTE data was included in the spectral analysis until MJD~52489. After this
date, the statistical quality of the HEXTE data was poor and was not included
in the analysis. We are presenting the first analysis of the HEXTE data for 
this source.

For all the observations, our first spectral model consisted of
absorption (``phabs'' in XSPEC), smeared edge
\citep[``smedge'' in XSPEC,][]{Ebisawa94}, a multicolor disk blackbody
\citep[``diskbb'' in XSPEC,][]{Makishima86}, a power law (``pegpower''
in XSPEC), and a narrow Gaussian to model the iron line. This model has been
commonly used for the spectral analysis of GBHs in the LH state
\citep{Tomsick00,Sobczak00}. A difference between our analysis and that
of \cite{Park03} is the modeling of the iron line. For most of our
observations, the statistical quality of the data is not sufficient for
reliable iron line studies, and including a narrow line instead of a broadened
Laor model iron line \citep{Laor91} yielded acceptable fits. For consistency
and simplicity, we left the iron line narrow, and we did not deduce any
physical results from the iron line fits. The hydrogen column density was
fixed to $\rm 4 \times 10^{21} \, cm^{-2}$, as used by \cite{Park03}.
The smeared edge width was fixed to 7 keV. Once we fit the observations with
this model, we added a high energy cut-off (``highecut'' in XSPEC) to the
model and refit. We included a high energy cut-off in the overall model if
the F-test indicates that adding this component significantly improves the fit.

At very low flux levels, the Galactic ridge emission becomes important.
Although \SO\ is not very close to the plane ($b$=+5.43$^{\circ}$), there was
some contribution from the ridge, evident by the detection of a narrow iron
line at 6.7 keV. However, even during the observations when \SO\ was at its
faintest (after MJD 52500), it was still detectable by the PCA. We obtained a
scanning observations on MJD 52547 that showed an increase in the
count rate when the \rxte\ pointing position reached the \SO\ position.
Therefore, we were not able to use any of our observations as background for
Galactic ridge emission. Instead, we model the ridge emission using the
description given in \cite{Revnivtsev04}. We also utilized the
\emph{XMM-Newton} observation on MJD~52504.54 that is close to our faint
\rxte\ observations \citep{Miller03_atel}. We fit our PCA observation close to 
the \emph{XMM-Newton} observation with a model consisting of
 interstellar absorption, a power-law to represent the \SO\ with
\emph{XMM-Newton} parameters, a second power-law and a narrow Gaussian to 
represent the ridge emission. We fixed the second power-law index to 2.15 
consistent with \cite{Revnivtsev04}, which resulted in
$\rm 1.2 \times 10^{-11} \; erg \; cm^{-2} \;s^{-1}$ Galactic ridge
contribution in the 3--25 keV band. If the index is not fixed, it results in
harder ridge emission.

\subsubsection{X-ray temporal analysis}\label{subsubsec:rxtetim}

For each observation, we computed the power density spectra (PDS)
and the cross spectra from the PCA data using IDL programs
developed at the University of T\"{u}bingen
\citep{Pottschmidt02th} for three energy bands, 3 -- 6 keV, 6 --
15 keV, and 15 -- 30 keV. We also computed PDS for the combined
band of 3 -- 30 keV. Above 30 keV, the source is not significantly
above the background for timing analysis. The PDS was normalized
as described in \cite{Miyamoto89} and corrected for the dead-time
effects according to \cite{Zhang95} with a dead-time of $\rm
10\,\mu s$ per event. Using 256 second time segments, we
investigated the low frequency QPOs and the timing properties of
the continuum up to 256 Hz. We fit all our PDSs with Lorentzians
of the form:
\begin{equation}
L_{i}(f)\;=\;{{R_{i}^{2} \; \Delta_{i}}\over{2 \, \pi \; [(f-f_{i})^{2}+({1\over{2}}\,\Delta_{i})^2]}}
\label{eq:lor}
\end{equation}
where subscript $i$ denotes each Lorentzian component in the fit, $R_{i}$ is
the rms amplitude of the Lorentzian in the frequency band of -$\infty$ to
+$\infty$, $\Delta_{i}$ is the
full-width-half-maximum, and $f_{i}$ is the resonance frequency. A useful
quantity of the Lorentzian is the ``peak frequency'' at which the Lorentzian
 contributes maximum power per logarithmic frequency interval:
\begin{equation}
\nu_{i}\;=\;{f_{i}\;\left({{\Delta_{i}^{2}}\over{4 \, f_{i}^{2}}}+1\right)^{1/2}}
\end{equation}

\begin{figure}[t]
\plotone{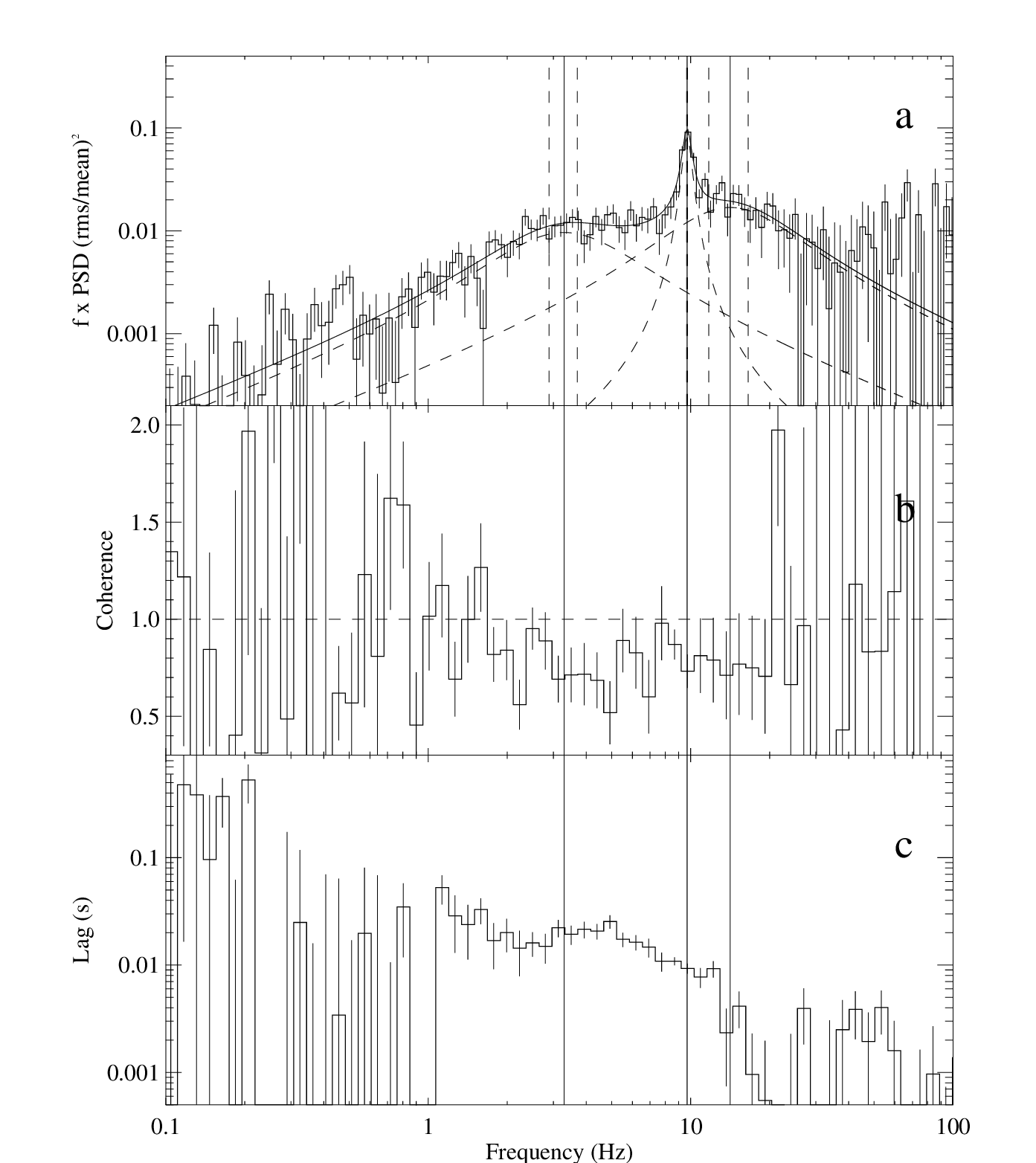}
\caption{\label{fig:lorex}
a. The power spectrum of Obs.9A in the 3--6 keV band, and in the form of 
PDS $\times$ frequency. The Lorentzians shown with dashed lines are individual 
components producing the overall fit shown with solid curve. The solid 
vertical lines are peak frequencies, and the dashed lines show the 1$\sigma$ 
errors on the peak frequencies. b. The coherence function for the same 
observation between the 3--6 keV band and the 6--15 keV band. c. The Fourier 
lag between the 3--6 keV and the 6--15 keV band for the same observation.
}
\end{figure}

An example power spectrum of \SO\ in the form of {PDS $\times$ frequency}
is shown in Fig.~\ref{fig:lorex}.a. The PDS consists of broad and narrow
Lorentzian fit components. In this figure, the Lorentzians peak at $\nu_{i}$,
demonstrating the easy identification of characteristic frequencies as peak
frequencies of Lorentzian components. The peak frequencies are shown with
solid vertical lines in Fig.~\ref{fig:lorex}.a. A Lorentzian with quality
value $Q_{i} = f_{i}/{\Delta_{i}}>2$ is denoted as a QPO (like the narrow 
feature in the middle in Fig.~\ref{fig:lorex}). The rms amplitudes are 
calculated over a frequency band from zero to infinity.

The cross spectrum leading to coherence function and time lag
constitutes another tool that is available as part of Fourier
analysis. The coherence function is a Fourier-frequency-dependent
measure of the degree of linear correlation between two concurrent
light curves measured simultaneously in two energy bands
\citep{Nowak99}. The Fourier time lag is a Fourier-frequency-dependent measure
of the time delay between two concurrent time series
\citep{Miyamoto89,Nowak99}. It is related to the phase of the average cross
power spectrum between the soft and hard band light curve. We use
 the convention that the sign of the lag is positive when hard photons lag soft
 photons. Observations of hard lags in GBHs have often been interpreted as
evidence for Compton upscattering in a hot electron gas \citep{Payne80},
however simple Comptonization models have difficulty explaining the magnitude
of lags \citep{Ford99}.

We calculated the coherence and lag spectrum between 3-6 keV and
6-15 keV band light curves for all observations. Not all
observations yielded meaningful coherence measurements; and the
lag measurements are only meaningful when the coherence is well
defined \citep{Nowak99}. Fig.~\ref{fig:lorex} shows the coherence spectrum
in the middle panel (b) and the lag spectrum in the bottom panel (c) for Obs.
9A. For observations close to the transition (such as the one in
Fig.~\ref{fig:lorex}), meaningful (not noisy) coherence and lag values were
observed between 1--10 Hz. To understand the trends in the amplitude
of these quantities, we calculated the mean lag and the mean
coherence, similar to what was done in \cite{Kalemci02}.

\subsection{Optical, and Infrared observations}\label{subsec:opir}

The OIR datasets were taken directly from \cite{Buxton04},
therefore we give a short summary of how these observations were
obtained. The details can be found in \cite{Buxton04}. Daily $V$-
and $J$- band images were taken using the YALO 1.0 m telescope from
MJD~52423. Daily $K$ band observations were initiated on MJD~52440.
The optical observations with $B$, $V$, and $I$ filters were conducted
between MJD~52442.9 and MJD~52500.8 using the 74 inch telescope at
Mount Stromblo Observatory. In this work, only the $J$
 band light curve is shown, as the important features are most visible in the
infrared. The light curves of the remaining bands can be found in
\cite{Buxton04}.

\subsection{Radio observations}\label{subsec:radio}

 The source was also observed at radio frequencies, and the details of MOST
and the Giant Metrewave Radio Telescope (GMRT) observations are discussed in
\cite{Park03}. Here, we give a summary of these observations. The source was
detected in radio several times between MJD~52443 and MJD~52447 by MOST and
GMRT. After MJD~52447, the source was radio quiet until MJD~52487. The upper
limits for MOST and GMRT observations on MJD~52480 are 3.0 mJy and 3.2 mJy
respectively. On MJD~52487, in the LH state, MOST detected the source at
$\rm 5.2 \pm 0.9$ mJy. The final observation by MOST on MJD 52496.33 did not
detect the source with an upper limit of 2.4 mJy \citep{Park03}.

We observed 4U 1543$-$47 five times during 2002 June and August with ATCA. In
each case the primary calibrator was PKS~1934$-$638 and the secondary
calibrator was PMN~J1603-4904. Observations took place at 4.80 and 8.64 GHz
with a bandwidth of 128 MHz. The data were reduced in the standard way with
(minimal) flagging, flux and phase calibration and finally mapping using
{\sc miriad}. A point source was fitted to the detected emission and the flux
density measured. A radio source was detected twice, firstly on MJD~52445 at
3.18$\pm$0.19 mJy (4.80 GHz), 2.76$\pm$0.07 mJy (8.64 GHz) and secondly on
MJD~52490 at 4.00$\pm$0.05 mJy (4.80 GHz), 4.19$\pm$0.06 mJy (8.64 GHz).
On MJD~52450, MJD~52451, and MJD~52453, during the TD state, the source was
radio quiet with upper limits ranging between 0.2 and 3.0 mJy.


\section{Results}\label{sec:results}

\subsection{Evolution in the X-ray regime}\label{subsec:evol}

\begin{figure}[t]
\plotone{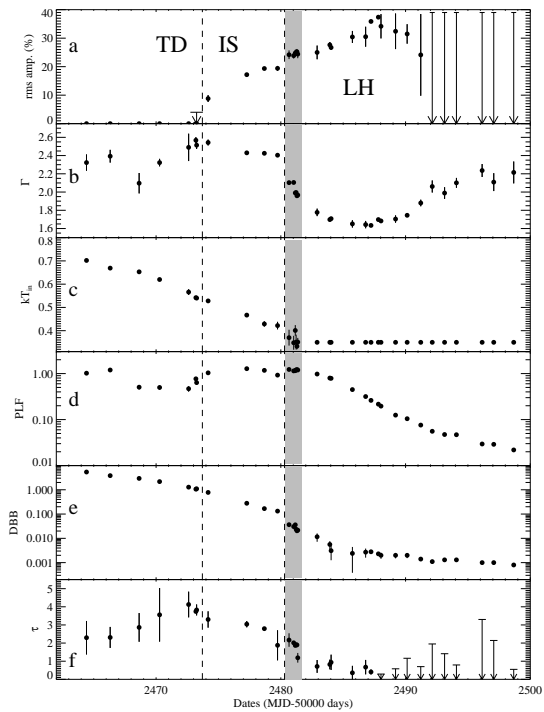} 
\caption{\label{fig:evol1}
The evolution of (a) the total rms amplitude of variability in
3-30 keV band, (b) the photon index (\IN), (c) the inner disk
temperature k\tin\, (d) the power-law flux in 3--25 keV band in
units of $\rm 10^{-9} \; ergs \; cm^{-2} \; s^{-1}$, (e) the
disk-blackbody flux in 3--25 keV band in units of $\rm 10^{-9} \,
ergs \, cm^{-2} \, s^{-1}$, (f) optical depth of the smedge
component in the fit. Unabsorbed fluxes were used. For most of the
flux measurements, the 1 $\sigma$ uncertainties are smaller than
plot symbols. 2 $\sigma$ upper limits are shown with arrows. The
dashed lines indicate approximate times of state transitions. The
gray area shows the observations during the transition to the LH
state. }
\end{figure}

To be able to determine the sequence of events during the decay of the
outburst, we investigated the evolution of several spectral and temporal fit
parameters as well as the IR and the radio fluxes. To establish the time of
state transition(s), we plotted several parameters as a function of time in
Fig.~\ref{fig:evol1}. The main spectral fit parameters are also tabulated in
Table~\ref{table:spe_par}.

\subsubsection{States and transitions}\label{subsubsec:states}

In Fig.~\ref{fig:evol1}, we marked the time of the first state transition
on \wsim MJD~52474 (between observations 7A and 8A in
Table~\ref{table:spe_par}), when the source showed a sharp increase in the rms
amplitude of variability, accompanied by a sharp increase in the power-law
flux.  Before this date, the spectra were dominated by the disk component
(more than 80\% of the total flux in 3--25 keV band), and hence the source was 
in the TD state. Within a few days, the power-law became the dominant 
component in the spectrum, and the rms amplitude of variability increased to 
\wsim\ 18\%. A QPO appeared in the PDS (see Fig.~\ref{fig:qpoind}). Note that 
during this time, while the rms amplitude of variability and the power-law 
flux were changing rapidly, the photon index remained almost constant, and 
the inner disk temperature (\tin) and the disk-blackbody (DBB) flux decayed 
smoothly, consistent with the observations reported in \cite{Kalemci03}. The 
source was in the IS during this time based on the softness of the photon 
index and the relatively comparable contributions of the hard and the soft 
components to the spectrum.

Around MJD~52480 (Obs.12B in Table~\ref{table:spe_par}), another set of sharp
 changes occurred; the rms amplitude of variability jumped to \wsim 24\% level,
 accompanied by a sharp hardening of the photon index, and rapid cooling of
the inner disk temperature. The power-law flux increased slightly,
and the DBB flux decreased, still smoothly, but more rapidly. The
overall effect was total dominance of the power-law flux over the
whole spectrum. After all these rapid changes (shown by the gray
area in Fig.~\ref{fig:evol1}), the source was in the LH state. In
this state, the rms amplitude of variability continued to
increase, before leveling off at \wsim 35\%. The photon index
stayed around 1.7, and then increased slightly after MJD~52490.
The power-law flux decayed smoothly. We fixed \tin\ to 0.35 as the
fit did not produce meaningful errors when we let \tin\ vary. The
disk component was only  affecting the first spectral bin in our
fit. As long as the $N_{\rm H}$ value was fixed, the DBB component was
required by the F-test in the fit, although its relative
contribution to the overall flux is less than 5 \% (see
Table~\ref{table:spe_par}).

The evolution of the smeared edge component in the spectral fits is also worth
 mentioning (see Fig.~\ref{fig:evol1}.f for the evolution of the optical 
depth). This component was required in the fit until MJD~52488, although its 
effects are less and less pronounced after MJD~52481. For observation 14C.d, 
the optical depth (which may be related to reflection fraction) in the smedge 
component was \wsim 2. It dropped down to \wsim 0.8 on MJD~52483, and stayed 
between 0.5--0.7 for the next six observations. This evolution may indicate 
that the reflection fraction dropped as the source went into the LH state. 
After MJD~52488, the smedge component was not required in the fit, although 
the reduced quality of the spectra did not permit us to place strong upper
limits on the optical depth of the component after MJD~52492.

\vspace{15pt}

\subsubsection{Short time scale evolution}\label{subsubsec:shortevol}

\begin{figure}[t]
\epsscale{1}
\plotone{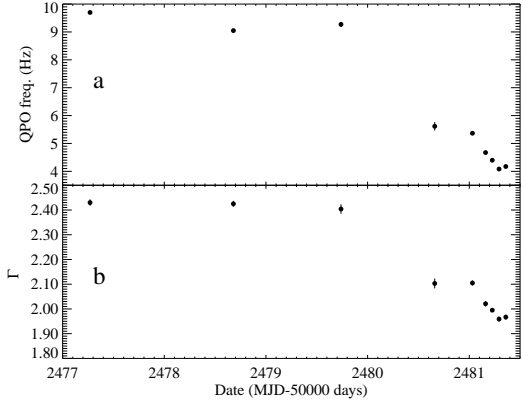}
\caption{\label{fig:qpoind}
(a) the QPO frequency (for the times when the QPO is observed), (b) the
photon index of 4U~1543$-$47 in the IS and during the transition to the LH
state.
}
\end{figure}

The first long observation of the program P70128 was well-timed, such that it
 was conducted when the source was in transition to the LH state while changes
were happening rapidly. In order to quantify these changes, we divided the
long observation into five sections (naturally separated by occultations, with
each piece having three PCUs on). Including these pieces, we show that the QPO
frequency and the photon index has a strong correlation (with linear
correlation coefficient of 0.995) in the IS and during the transition
(see Fig.~\ref{fig:qpoind}). Note that the last six observations are only 2-3
hours apart! The QPO frequency - photon index correlation has been known
\citep{Kalemci_tez,Vignarca03}, but for the first time, we show that it holds
on timescales as short as hours.

\begin{table}[hb]
\caption{\label{table:cutoff} High energy cut-off parameters}
\begin{minipage}{\linewidth}
\renewcommand{\thefootnote}{\thempfootnote}
\scriptsize
\begin{tabular}{l|c|c|c} \hline \hline
Obs. \# & Date & $E_{cut}$ (keV) & $E_{folding}$ (keV) \\ \hline
13C & 52481.03 & $\rm 56 \pm 15$ & $\rm 98 \pm 53$ \\
14C.a & 52481.16 & $\rm 56 \pm 10$ & $\rm 79 \pm 32$ \\
14C.b & 52481.23 & $\rm 38 \pm 15$ & $\rm 197 \pm 88$ \\
14C.c & 52481.29  & $\rm 45 \pm 7$ & $\rm 86 \pm 28$ \\
14C.d & 52481.36  & $\rm 34 \pm 13$ & $\rm 318^{+411}_{-127}$\\
\end{tabular}
\end{minipage}
\end{table}

Another very important result that came out from the long
observation is the presence of a high energy cut-off in the HEXTE
spectrum during the state transition. The cut-off is only
detected significantly during the observations for which the
power-law index was changing rapidly, and the transition was
taking place (gray area in Fig.~\ref{fig:evol1}). The cut-off and
the folding energies for those observations are given in
Table~\ref{table:cutoff}. Errors are too large to establish a
pattern in the folding energy, except for observation 14C.d for
which the folding energy increased. This increase is probably a
sign of going back to regular power-law shape. After this
observation, the fit values for the folding energy are far beyond the
\hexte\ energy range, and a cut-off component is not required by
the F-test. This is not the first time that the existence of a
high energy cut-off has been observed during a state transition.
XTE~J1550$-$564 during its decay in 2000 outburst shows similar
behavior: the cut-off was only significant during the transition
\citep{Tomsick01b, Rodriguez03}.

\subsubsection{Evolution in temporal parameters}\label{subsubsec:tempevol}


\begin{figure}[t]
\plotone{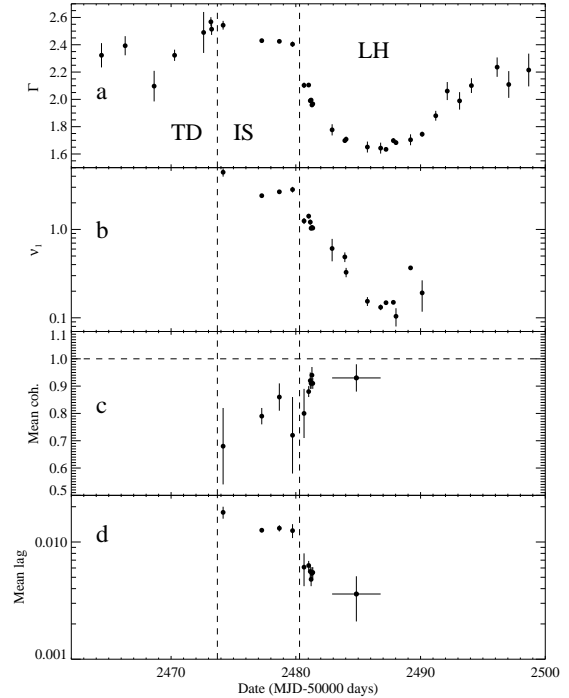} 
\caption{\label{fig:lagcoh}
Evolution of: (a) the photon index (same as
Fig.~\ref{fig:evol1}.b), (b) the lowest peak frequency
($\nu_{1}$), (c) the mean coherence in the 1--10 Hz band between the 3--6
keV and the 6--15 keV band, and (d) the mean lag in the 1--10 Hz band
between the 3--6 keV and the 6--15 keV band. The dashed lines represent
approximate times of state transitions. Five LH state observations were
merged to obtain the last point.}
\end{figure}

The evolutions of the total rms amplitude of variability and the QPO 
frequencies are shown in Figs.~\ref{fig:evol1}~and~\ref{fig:qpoind}. The QPO 
has been detected only for the IS and during the transition to the LH state. In
 the LH state, the PDS can usually be characterized with 2 broad Lorentzians.
Therefore, to understand the evolution of characteristic frequencies, we used
 the broad Lorentzian with the lowest peak frequency that is present for all
observations ($\nu_{1}$, see Equation 2). This evolution is shown in
 Fig.~\ref{fig:lagcoh}.b. In this figure, we also show the evolution of the
photon index (Fig.~\ref{fig:lagcoh}.a), which is known to correlate with the
temporal parameters that are discussed in this section. There is a good
correlation between the photon index and $\nu_{1}$, which is also observed
for other sources \citep{Kalemci_tez,Pottschmidt03}. The decrease of
characteristic frequencies in time after the transition to the LH is also a
known effect \citep{Kalemci03}. Note that the characteristic frequency levels
off as the photon index also levels off at low flux levels.

We also included the evolution of mean coherence and lags in 1--10
Hz band in Fig.~\ref{fig:lagcoh}. It was necessary to merge the
light curves of the 5 observations in the LH state (15B through
19B in Table~\ref{table:spe_par}) to improve statistics. In the
IS, the coherence is relatively low, whereas the lags are high. As
the source enters the LH state, the coherence increases and
approaches unity, and the lag decreases. In the LH state, the
coherence is high, and the mean lag is low. These results are in
agreement with the analysis of other black hole transients; high
lag, low coherence during the IS, and high coherence, low (usually
consistent with zero) lag in the LH state, along with a
correlation between photon index and the 1--10 Hz mean lag when
they are measurable \citep{Kalemci_tez}.


\begin{figure}[t]
\plotone{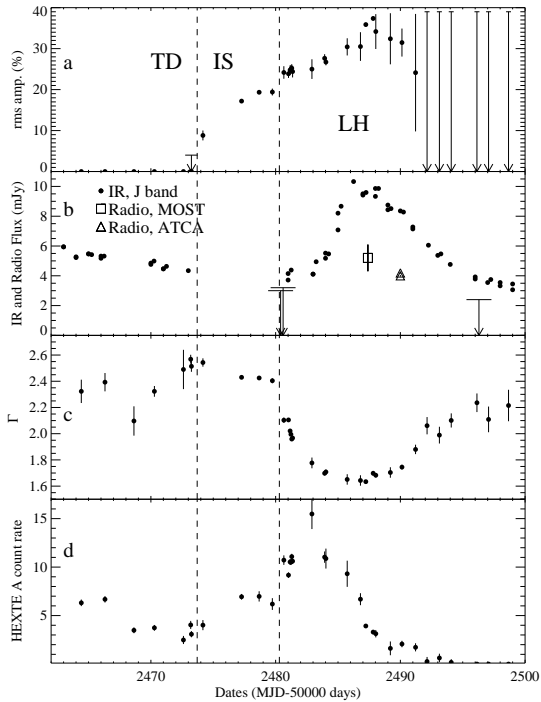}
\caption{\label{fig:evol2}
The evolution of a. the rms amplitude of variability, b. the $J$ band infrared
fluxes from \citealt{Buxton04} (dots) along with radio fluxes from our
observations (triangles) and \citealt{Park03} (upper limits and the square),
c. the photon index, and d. HEXTE Cluster A count rate in 16-100 keV band. Note
that panels a and c were shown before in Fig.~\ref{fig:evol1}. The gap in the
IR light curve during the IS is due to lack of coverage.
}
\end{figure}

\subsection{Multi-wavelength observations}\label{subsec:multi}

Although some of the radio, optical and IR data from this source were
presented before \citep{Park03,Buxton04}, additional information is available
when they are combined with the data from pointed \rxte\ observations. The
evolution of the $J$ band infrared fluxes, along with the radio information is
shown in Fig~\ref{fig:evol2}.b. In this figure, we also show the rms amplitude
of variability, the photon index (same as Fig.~\ref{fig:evol1}.a, b
respectively), and also the HEXTE Cluster A count rate in 16--100 keV band.
The $J$ band flux decreases steadily in the TD state as the source flux was
decaying. Unfortunately, there was no coverage during the first transition,
making it impossible for us to determine if there was a response in the
OIR emission. On the other hand, there was strong response after
the transition to the LH state such that the $J$ band flux started to increase
 sharply \wsim3 days after the time of transition, and peaked in 3 days. A
two-sided Gaussian fit to the $J$ band light curve during the secondary maximum
results in a peak time of MJD~$\rm 52486.3^{+1.5}_{-1.8}$ \citep{Buxton04}.
After staying at the peak for a few days, it started to decay, and reached the
 pre-secondary peak levels in 7 days. As shown in \cite{Buxton04}, both the
optical ($B$, $V$, $I$) and the infrared light curves show this peak. The peak 
is most pronounced in the $J$ and the $K$ bands as the strength of 
emission increases with wavelength. The spectrum of the OIR points is best fit 
by a power-law rather than a blackbody, or a disk-blackbody model 
\citep{Buxton04}. This strongly suggests that the origin of the OIR emission 
is optically thin synchrotron. From the beginning of our observations 
(MJD~52464) to MJD~52487, there was only one radio observation that resulted
 in no detection. The detections occurred in the LH state, during the secondary
 maximum. The radio emission most likely originates from an outflow, 
possibly a compact jet based on the inverted spectrum of the observation 
24B (see Fig.~\ref{fig:sed2}).

While the $J$ band flux increased, the rms amplitude of variability
(Fig.~\ref{fig:evol2}.a)  also increased slightly. It is not clear
what happened to the rms amplitude of variability after MJD~52492,
as the count rate was too low to constrain the timing properties. The photon 
index (Fig.~\ref{fig:evol2}.c), on the other hand, seemed to show an
anti-correlation with the $J$ band flux. There is a \wsim 2 day time
lag between the time where the photon index began to harden and
the $J$ band flux began to increase. As the $J$ band flux decreased,
the photon index began to soften. \footnote{When not fixed in
the fit, indices of the Galactic ridge emission were consistently
harder than 2.15 for the last two points. We fixed the ridge
power-law index to 2.15 so the fit resulted in a harder index for
the source. The actual photon indices for these two points were
probably higher.}

Another interesting change with respect to the changes in the
OIR is the evolution of the hard flux. This is represented by the
HEXTE cluster A count rate in 16--100 keV band in
Fig.~\ref{fig:evol2}.d. First of all, there was a sharp increase
in the count rate right at the transition (probably caused by
sharp hardening of the X-ray spectrum). By the time HEXTE count
rate peaked, the power-law index dipped, and the $J$ band flux
started to rise. The lag between the time that hard X-rays peaked
and the OIR peaked was about 2 days.

\subsection{Spectral Energy Distribution}\label{subsec:sed}

\begin{figure}[t]
\plotone{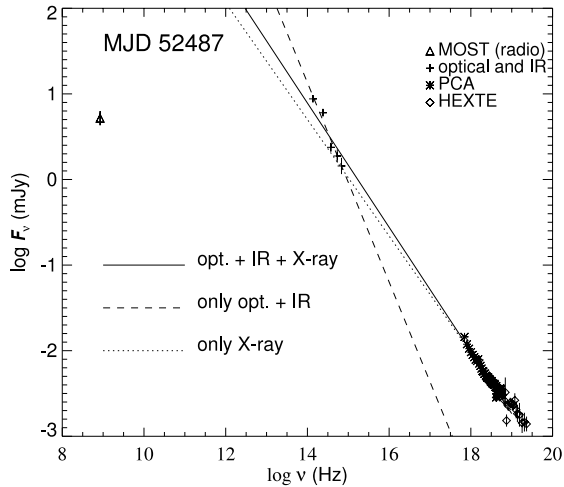}
\caption{\label{fig:sed1}
Spectral energy distribution of \SO\ at MJD~52487. The lines represent power
law fits to different subsets of data (see Table.~\ref{table:sedfits}).
}
\end{figure}

\begin{figure}[t]
\plotone{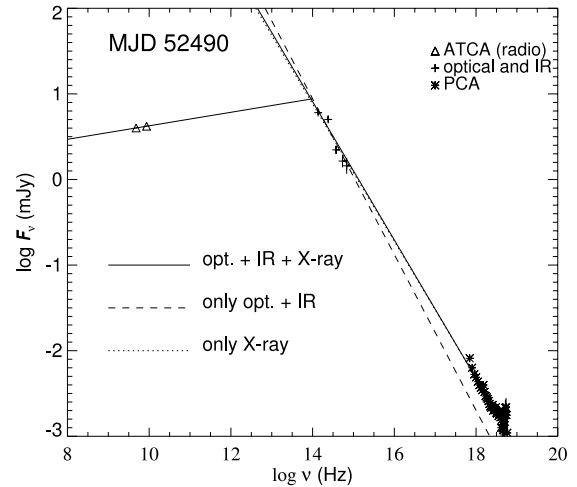}
\caption{\label{fig:sed2}
Spectral energy distribution of \SO\ at MJD~52490. The lines passing through
the optical, IR and X-ray points represent power law fits to different subsets
of data (see Table.~\ref{table:sedfits}). Another line is drawn that passes
through the radio points and intersects the power-law fits.
}
\end{figure}

We have constructed two spectral energy distributions (SED) close
to the peak of the IR maximum for the dates of the MOST (\wsim
MJD~52487) and ATCA (\wsim MJD~52490) observations. We first
corrected the OIR magnitudes for interstellar extinction using E($B$
-- $V$) = 0.50 $\pm$ 0.05 \citep{Orosz98}, $\rm A_V$ = 3.2 E($B$ -- $V$)
= 1.60 $\pm$ 0.16 \citep{Zombeck90}, $\rm A_{B}/A_{V}$ = 1.324,
$\rm A_{I}/A_{V}$ = 0.482, $\rm A_{J}/A_{V}$ = 0.282, $\rm
A_{K}/A_{V}$ = 0.112 \citep{Rieke85}. There were no $B$, $V$, or $I$
observations on MJD~52487 \citep{Buxton04}. The source was observed in $B$, 
$V$, and $I$ bands on MJD~52486, and again on MJD~52490. We used linear
interpolation to estimate the peak fluxes on MJD~52487. For each band, we
fitted a line to the underlying magnitudes using the points before and after
the peak to estimate the underlying flux during the secondary maximum. We 
converted the peak and underlying magnitudes into fluxes using the irradiation
factors and description in \cite{Zombeck90}, and then subtracted 
the underlying flux from the peak flux to obtain the residual flux
representing the emission from the peak only \citep[similar to
what was done in][]{Buxton04}. We used a measurement error of 0.05
magnitudes in each band. In addition, we calculated the error due
to uncertainty in extinction using the uncertainty in E($B$ -- $V$).
We also obtained 0.05 and 0.08 magnitude uncertainties in determining the
baseline fluxes for $J$ an $K$ bands, and for $B$, $V$, and $I$ bands, 
respectively, from the fitting process. For MJD~52487, an additional 0.05 
magnitude uncertainty is estimated in $B$, $V$, and $I$ bands due to 
interpolation in finding the peak flux. All the error components were added in 
quadrature to estimate the total uncertainty in the measurements. The 
unabsorbed PCA and HEXTE fluxes for MJD~52487, and the unabsorbed PCA fluxes 
for MJD~52490 are converted into mJy to complete the SEDs shown in 
Figs~\ref{fig:sed1}~and~\ref{fig:sed2}.


 For each case, we fit a power law to the three sets of data; OIR
set only, X-ray set only, and OIR and X-ray (overall) set.
Fig.~\ref{fig:sed1} shows the SED at MJD~52487 which includes
the MOST observation, optical and IR data, as well as PCA and HEXTE data, and
 the power-law fits (see Table.~\ref{table:sedfits} for fit parameters).
Statistically, the OIR points are not the continuation of the power-law
of the X-ray data as the high quality PCA+HEXTE data has a photon index of
-1.63 with reduced $\chi^{2}$ value of 0.64 (XSPEC fit, including all other
model components). The optical and IR points alone result in a much steeper
power law index of 2.16. 

\begin{table}[h]
\caption{\label{table:sedfits} SED power-law fit parameters}
\begin{minipage}{\linewidth}
\small
\begin{tabular}{c|c|c} \hline \hline
\multicolumn{3}{c}{MJD~52487} \\ \hline
Data set  &  photon index  & $\chi^{2}$/DOF  \\
Opt. and IR only & $\rm -2.16 \pm 0.10$ & 8.74 / 3  \\
X-ray only\footnote{X-ray only fits for both dates are from XSPEC fits, with all other model components} & $\rm -1.63 \pm 0.01$ & 39.78 / 62 \\
Opt., IR, and X-ray  & $\rm -1.73 \pm 0.01$ & 164.30 / 64 \\ \hline
\multicolumn{3}{c}{MJD~52490} \\ \hline
Opt. and IR only & $\rm -1.91 \pm 0.10$ & 14.66 / 3  \\
X-ray only & $\rm -1.75 \pm 0.02$ & 47.68 / 43  \\
Opt., IR, and X-ray & $\rm -1.80 \pm 0.02$ & 80.39 / 42  \\ \hline
\end{tabular}
\end{minipage}
\end{table}

Although the OIR points do not follow the power-law fit of the
X-ray data again for the SED of MJD~52490 shown in
Fig.~\ref{fig:sed2}, the discrepancy is much less compared to
MJD~52487 observation. The X-ray and OIR power-law indices are
consistent within 2-$\sigma$ uncertainty. Unlike the first case,
this SED provides the break point of the spectrum as there are two
radio points. The turnover is around $10^{14}$ Hz. This value is
similar to the observed values of other sources
\citep{Corbel02,Homan04}.

For the B, V, and the I bands, the largest contribution to the error estimate 
is from the reddening factor. We investigated whether an underestimated error
 in the reddening could cause the discrepancy between the OIR slope and X-ray 
slope for our SEDs. For the SED on MJD~52490, an  uncertainty of 15\% in 
$\rm A_{V}$ instead of the quoted 10\% is enough to make the OIR slope and the 
X-ray slope be consistent within 1$\sigma$ uncertainty. However, for the SED 
on MJD~52487, the uncertainty of the reddening factor must be as high as 60\% 
to make OIR slope and X-ray slope be consistent, which is highly unlikely.

\subsection{Summary of results}

In this section we summarize our main results that will be discussed in detail
in \S~\ref{sec:discussion}.

\begin{itemize}

\item \SO\ showed two state transitions during its decay in the 2002 outburst:

{\subitem -the first transition was from the TD state to the IS on
\wsim MJD~52474, and was marked by a jump in rms amplitude of variability and 
power-law flux.}

{\subitem -the second transition was from the IS to the LH state, and was
 marked by a sharp hardening of the X-ray spectrum, and an increase in rms 
amplitude of variability.}

\item {During the IS and the transition to the LH state a QPO was
detected. The resonance frequency (see Eq.~\ref{eq:lor}) of the QPO showed a 
very strong correlation with the photon index on time scales of hours.}

\item A high energy cut-off was detected in the X-ray spectrum only during the
transition to the LH state, whereas after the transition and during the LH
state, the cut-off was either absent or beyond the HEXTE detection range.

\item The spectrum softened at very low flux levels in the LH state.

\item The characteristic frequencies decreased during the transition and in the
LH state before they leveled off. The characteristic frequencies and the
photon index showed a good correlation.

\item The mean coherence was low during the IS, and then increased and 
approached unity as the source made the transition to the LH state. The
mean lag, on the other hand, was high in the IS, and decreased
during the transition.

\item The OIR light curves started peaking \wsim 3 days after the transition to
the LH state. The radio detections occurred during the OIR secondary maximum.

\item The HEXTE 16-100 keV light curve increased during the transition to the
LH state while the photon index hardened. It did decay, however, while the
IR flux was rising.

\item The SEDs show that the OIR part of the spectrum can be
represented by a power-law. For the SED on MJD~52487, the hard
X-ray points are not a continuation of the OIR points and have a
different index. For the SED on MJD~52490, the difference is less,
and the X-ray and OIR power-law indices are consistent within 2-$\sigma$ 
uncertainty.

\end{itemize}


\section{Discussion}\label{sec:discussion}

The multi-wavelength observations of the 2002 outburst of 
\SO\ have provided a wealth of information on the physical
processes during state transitions and the LH state of GBH
transients. In this section, using our observational results, previous
experience from other sources, and some theoretical models, we
try to place constraints on the X-ray emission geometry, the
hard X-ray emission mechanism (power-law part of the X-ray
spectrum), and we try to identify the origin of seed photons (if seed
photons are required). These three properties constitute the
fundamental differences between various models of X-ray emission
from GBHs \citep{Tomsick04b}.

One set of models considers a corona (which could be in a form of
an ADAF) inside a transition radius (that can move) as the
accretion geometry \citep[e.g, see][]{Esin97}. In this case, the
base of the jet may be the whole corona, or the base may be small
compared to the overall size of the corona
\citep{Fender01b,Markoff04}. We note that the jet has not been
incorporated into the ADAF or "sphere+disk" models in a self
consistent way. In the ``magnetic corona'' model, active,
optically thin regions above the disk
 are responsible for the hard X-ray emission and the outflow formation
\citep{Merloni02}. In this model, the position of the inner edge of the
accretion disk does not necessarily change significantly. For these cases where
 the main hard X-ray producing mechanism is Compton scattering, the seed
photons could either be produced in the accretion disk as
blackbody radiation or in the corona as synchrotron radiation
\citep[synchrotron self-Compton,][]{Markoff04}. On the other hand,
the main hard X-ray emission mechanism may not be Compton
scattering, but synchrotron radiation originating in the optically
thin shock region in the jet \citep{Falcke99,Markoff01,Markoff04}.
A combination of emission mechanisms is also possible \citep[see,
for example][]{Rodriguez04}, synchrotron radiation and Compton
scattering may contribute simultaneously, also, the seed photons
for Compton scattering may be a combination of synchrotron at the
base of a jet or accretion disk. However, unless these mechanisms
conspire to result in very similar observational properties, it
may still be possible to determine the major emission mechanism
and major contributor of seed photons.

 The transition of \SO\ to the IS from the TD state shows the characteristics
that many other sources also show, namely a strong jump in the rms amplitude of
 variability (from less than 4\% to $\sim$9\%) accompanied by an increase in
the power-law flux in a few days timescale. Note that during this transition,
the change in the power-law photon index is smooth. Based on similar evolution
of spectral and temporal parameters of other GBH transients, \cite{Kalemci03}
concluded that a threshold volume for a hot corona is required to observe
variability. The second transition during which the photon index
and the inner disk temperature decrease very rapidly, and the
multi-wavelength observations in the LH state provide more insight
into understanding the origin of hard X-ray emission mechanism for
this source. Below, we describe how the sequence of observational
changes may be explained by different hard X-ray emission
mechanisms and geometries.

\subsection{Compton scattering - recessing accretion disk}

In the IS, the inner edge of the disk is close to the black hole,
the high reflection fraction indicates that the disk and the
corona have some overlap, and there is strong cooling of coronal electrons
by the accretion disk seed photons. The electron distribution is more likely
not Maxwellian \citep{Coppi00}, as no cut-off is detected in the \hexte\
spectrum. It is possible that an outflow may already be present in this state
which provides an``elongated'' and variable corona structure at its base
which could explain large lags and small coherence values, as well as the
non-thermal nature of the electron population
\citep{Pottschmidt00,Kalemci02}.

At some critical combination of physical parameters \citep[mass
accretion rate has been shown not to be the only parameter
determining when the transition occurs,][]{Homan01,Kalemci03}, the
inner parts of the disk evaporate rapidly into a hot corona and
the transition radius moves outward. During this process, the
optical thickness of the corona increases and the thermal
electrons dominate the system, causing a harder spectrum with a
cut-off in the \hexte\ band. As the disk moves outward, the
hardening of the spectrum continues because of a lower level of
cooling by the disk (the cut-off becomes undetectable), the disk
temperature decreases as the hotter inner parts of the disk
evaporates. The recession of the disk could also explain the
evolution of the QPO frequency and $\nu_{1}$ if they are related
to some dynamical timescale of the inner disk. This model may
further explain the correlation between the photon index and the
QPO if the hardening is a direct result of less cooling by
increasing inner disk radius that the QPO frequency is linked.
Similar arguments had been put forward to explain the well
established photon index-reflection fraction correlation in GBHs
and AGNs \citep[][and references therein]{Zdziarski02_2}. The
relation between the photon index and the QPO frequency could
also be due to increasing optical depth of the corona as it
expands \citep{Titarchuk04}.

The detection of the inverted radio spectrum, and also the IR peak
indicate that a compact jet is present deep in the LH state. The
high coherence and low lag values, and the indication of low
reflection fraction may point out that the size of the jet is now
small compared to the transition radius. According to
\cite{Meier01_b}, the production region for a MHD driven compact
jet is 7 - 8 $R_{g}$ for a Schwarzchild hole, and even smaller for
a Kerr hole. More than one order of magnitude decrease in
characteristic frequency between the IS and the deep LH state may
also indicate that the transition radius is away from the jet
production region. The similarity of the PDS in the IS with the
detection of the QPOs, and the photon index  - QPO frequency
correlation point to accretion disk as the major source of seed
photons in the IS and during the transition. Also, once the
spectrum reaches its hardest point on \wsim MJD~52483, the
HEXTE count rate starts to decrease, in parallel with the decrease
in the DBB flux, again indicating that the major source of seed
photons is the accretion disk. However, the source of seed photons
deep in the LH state is unclear, and the synchrotron-self Compton
may be contributing to the hard X-ray emission after \wsim
MJD~52487. The spectral softening at the end of the outburst may
be due to a decrease of the optical thickness of the Comptonizing
medium as the mass accretion rate decreases, and the
 energetic electrons either advectively accrete onto the black hole, or leave
the vicinity of the black hole in the form of an outflow \citep{Esin97}. We
note that this type of softening is observed in many sources
\citep{Kalemci_tez, Tomsick01b}.

\subsection{Compton scattering - stable inner disk (magnetic
flares):}

 The fits to the PCA spectrum, even at very low flux levels,
require a disk component to be able to fit the first bin. However,
we cannot reliably constrain disk parameters after MJD~52485 with
PCA observations when only one bin is affected. The
\emph{XMM-Newton} observation that took place on MJD~52504.5
indicates that a disk component is required in the fit at
$\sim$3 $\sigma$ level with a temperature of 0.19 keV. After
converting the normalization of the diskbb fit to an inner disk
radius, \cite{Miller03_atel} claimed that the disk may still be
very close to the last stable orbit. If the disk is not moving, as
in the magnetic flares model, the explanation of the spectral
evolution is different than the case of the moving disk. In the
stable disk case, after the transition to the IS, the active
regions increase in number compared to before the transition and
the variability is observed due to short lived flares in the
active regions \citep{Merloni01}. A non-Maxwellian distribution of
electrons in the corona is again required in the IS state. During
the transition to the LH state, a larger and larger portion of the
accretion energy is used to buoy the magnetic flux tubes and to
heat the corona. This results in cooling the disk, and hardening
the photon index. At lower mass accretion rates, the accretion
energy might be channeled into launching and sustaining the jet
rather than heating the electrons which may explain the softening
at the end of the outburst \citep{Merloni02}. The decrease in the 
reflection fraction could be explained if the surface of the
disk becomes ionized which washes away the features
\citep{Nayakshin01}. However, a recent study showed that the
magnetic flares model cannot explain the high energy emission of
Cyg X-1, even with the the ionization effect \citep{Barrio03}. It
is also not clear what determines the QPO frequency, what causes
the temporal evolution, why the characteristic frequencies
decrease in time, and what causes the QPO frequency - photon index
correlation.

\subsection{Direct synchrotron radiation}
It is hard to explain the sequence of events in this source if the
main hard X-ray emission mechanism is direct synchrotron from a
shock region (first acceleration zone) in a jet for most of the
outburst decay. The origin of the OIR peak emission is probably
synchrotron from a shock region in the jet (where the hard X-rays
would also originate) as indicated by SEDs \citep{Buxton04}. The
IR peak shows that the shock region has formed after the
transition to the LH state. Even though there is no coverage in
the OIR during the IS, the IR emission between MJD~52480 and
MJD~52483 is the continuation of the decay in IR before the IS,
and it is unlikely that another IR peak is present in the IS. The
outflow may be present after the first transition to the IS;
however, the formation of the shock region did not take place
before the source was in the LH state. If the assessment above is
correct, the synchrotron X-ray emission from the jet region could
not have been produced before the source is deep in the LH state.
But it is possible that the synchrotron emission becomes the
dominant X-ray emission mechanism after the formation of the shock
region. The synchrotron model results in very low reflection
fraction \citep[\wsim 3\%, ][]{Markoff04}. There is a sharp drop
in smeared edge optical depth just before the IR flux begins to
rise. It is also intriguing that the QPO disappears, the structure
of the power spectrum changes as the reflection decreases.
However, there is also evidence against the synchrotron
interpretation even after the IR flux begins to rise. The SEDs
indicate that, even without an exponential cut-off, the X-ray
contribution from synchrotron is below the observed levels. More
importantly, the \hexte\ light curve does not
 follow the IR light curve, decreasing while the IR is still
rising, and the photon index is hardening between MJD~52483 and MJD~52487
 (see Fig.~\ref{fig:evol2}). This is  very hard to explain in the synchrotron
model when the IR and optical emission comes from the optically thin part of
 the SED as in \SO\ (see Figs~\ref{fig:sed1}~and~\ref{fig:sed2}). We
emphasize that the synchrotron radiation from the jet may still be
contributing to the overall X-ray emission, but it cannot be the
dominant emission mechanism until \wsim MJD~52490. It is hard to
make a claim about the dominant emission mechanism after this
date. Note that, the discrepancy in the power-law index of the OIR
points and X-ray points on MJD~52490 is also much less than that
of MJD~52487 (see Figs.~\ref{fig:sed1}~and~\ref{fig:sed2}).

\subsection{Comparison with GX 339$-$4}
 When only evolution of X-ray properties are considered,
\SO\ is a very typical source showing characteristics of many
other GBH transients \citep{Kalemci03}. However, this does not
translate easily into generalizing the results from this work to
all GBH transients as few objects had $\sim$daily simultaneous IR and X-ray
coverage that one could compare the details of spectral evolution.
GX~339$-$4 had such coverage during its 2002 outburst.
It is also a very typical GBH transient in its spectral and timing properties.
 In fact, its 2002 outburst decay shows very similar spectral and temporal
characteristics in X-rays to the decay of \SO\ (Kalemci et al., in
preparation). However, in the LH state during the outburst rise,
GX~339$-$4 shows a correlation between the X-ray flux and the OIR
flux \citep{Homan04}, whereas \SO\ shows an anti-correlation in
the LH state during the decay. Although somewhat speculative, the
main difference between the behavior in \SO\ and in GX~339$-$4
could be the origin of seed photons for Compton upscattering. For
\SO, the decay part of the HEXTE light curve seems to have two
slopes, a sharp decay until MJD~52486, and a more gradual decay
after this date. This could be a sign of the major seed photon
emission mechanism changing from DBB from the accretion disk to
synchrotron self-Compton at the base of the jet after MJD~52489 as the IR and
X-rays correlate after this date. Complete multi-wavelength analysis
(including OIR data) of other black hole transients is necessary
to generalize the relation between the OIR and X-ray flux.

\section{Summary and Conclusion}

Using X-rays, OIR, and radio observations, we characterized the
outburst decay of the GBHT \SO, and placed constraints on several
emission models. A large, non-thermal, and radiatively inefficient
outflow could explain the spectral and temporal evolution in the
IS. The presence of high energy cut-off in the X-ray spectra
during the transition to the LH state is indicative of a thermal
electron distribution. In general, our observations are consistent
with a recessing accretion disk + hot corona + ``compact'' jet
geometry with the main hard X-ray emission mechanism of Compton
upscattering of soft accretion disk seed photons by energetic
electrons in the corona. Here we used the word ``compact'' to also
emphasize the size of the base of the jet, which most likely is
small compared to the overall size of the corona. This
interpretation can reasonably explain all aspects of our
observations until deep in the LH state.

Our results do not strongly rule out the possibility of accretion
disk being always close to the last stable orbit as in magnetic
flares model; however, the QPO frequency - photon index
correlation, and the decrease of characteristic frequencies in
time are two results from this and other sources that need to be
understood for this model \citep{Zdziarski03,Tomsick04b}.

Our observations disfavor synchrotron from a shock region in the
jet as major source of hard X-rays until deep in the LH state. We
cannot place constraints on different emission models after MJD
52490.


\acknowledgments E.K. acknowledges NASA grant NAG5-13142 and
partial support of  T\"UB\.ITAK. The authors thank Sera Markoff,
Jeroen Homan, and Juri Poutanen for useful discussion and
comments. EK thanks all scientists who contributed to the
T\"{u}bingen Timing Tools. Authors thank Charles Bailyn who (along
with M. M. Buxton) provided access to the OIR data prior to
publication. J.A.T. acknowledges partial support from NASA grant
NAG5-13055. M. B. gratefully acknowledges support from the National Science
Foundation through grant AST-0098421. R.E.R acknowledges NASA grant NAS5-30270.
KP acknowledges support from the Deutsches Zentrum f\"ur Luft- und
Raumfahrt grant 50 OG 95030. The authors also would like to thank Jean Swank 
and Evan Smith for all their scheduling efforts, especially scheduling the 
first P70128 observation as early as possible, that observation could
not have taken place at a better time. We also would like to thank
Rob Fender for his help coordinating the radio observations, and Steven Tingay
for conducting the ATCA observations. The Australia Telescope Compact Array is 
part of the Australia Telescope which is funded by the Commonwealth of 
Australia for operation as a National Facility managed by CSIRO. This
work has made use of the HEASARC at GSFC.





\end{document}